# Phonons correlation with magnetic excitations in weak ferromagnet YCrO$_3$

Yogesh Sharma, Satyaprakash Sahoo, W. Perez and Ram S. Katiyar*

Department of Physics, University of Puerto Rico, PR-00936-8377, USA

**ABSTRACT:**

We report on the temperature dependent Raman spectroscopic studies of orthorombic distorted perovskite YCrO$_3$ in the temperature range of 20-300K. Temperature dependence of DC-magnetization measurement under field cooled and zero field cooled modes confirmed the transition temperature ($T_N$ ~142K) and anomalous characteristic temperature ($T^*$ ~60K), above which magnetization tends to saturate. Magnetization isotherm recorded below $T_N$ at 125K shows clear loop opening without magnetization saturation up to 20kOe, indicating the coexistence of antiferromagnetic (AFM) interaction with weak ferromagnetic (WFM) phase. Mean field calculation for exchange constants further confirms the complex magnetic phase below $T_N$. Temperature evolution of lineshape parameters of selected modes (associated with the octahedral rotation and A-shift in the unit cell) revealed anomalous phonon shift near Cr$^{3+}$ magnetic ordering temperature ($T_N$ ~142K). Additional phonon anomaly was identified at $T^*$ ~60K in agreement with the magnetization results and reflects the change in spin dynamics, plausibly due to the change in Cr-spin configuration. Moreover, the positive and negative shift in Raman frequency below $T_N$ revel the existence of competing WFM and AFM exchanges. The phonon shift of $B_{3g}$ (3)-octahedral rotation mode fairly scaled with the square of sublattice magnetization from $T_N$ to $T^*$, below which it start to depart from the conventional behaviour and need further attention. This correlation between magnetic and Raman data elucidate the spin-phonon coupling owing to the multiferroic phenomenon in YCrO$_3$.

Corresponding author: rkatiyar@hpcf.upr.edu

## 1. INTRODUCTION:

In the last decade, the rare earth chromites RCrO$_3$ (where R= La, Gd, Er, Ho , Yb, Lu and Y) have witnessed a renewed interest as multiferroic material due to the possibility of combining ferroelectric and ferromagnetic ordered parameters in a single phase.[1,2,3] RCrO$_3$ compound exhibit an orthorhombic pervoskite structure with space group Pnma. A rotation

(tilt) of rigid $CrO_6$-octahedra leads orthorhombic distortion, which can be continuously tuned by the ionic size of the rare-earth ion.[4] Recent finding by the Weber et al.[4] showed that the some of the $A_g$ Raman-active modes with octahedral-rotation vibrational pattern are an excellent signature for determining the magnitude of the octahedra rotation in $RCrO_3$. It is widely accepted that, this cooperative octahedral rotation is equivalent to the reduction in Cr-O-Cr bong angle, which in turn produces canted antiferromagnetic (CAFM) order of $Cr^{3+}$-spin. In $RCrO_3$, the weak ferromagnetism (WFM) resulting from a small canting of the $Cr^{3+}$-spins, and the $Cr^{3+}$-sublattice moment is observed to lie along the a-axis or the c-axis.[5] However, in some orthochromites, the coupling between the magnetic moment on R-ions and the $Cr^{3+}$-spins is responsible for an easy-axis rotation below Neel Temperature ($T_N$), from one crystal axis to another with varying temperature or applied magnetic field.[6] Therefore, some of the rare earth chromites have been gained considerable interest for their rich and exotic magnetic behaviour below $T_N$[5,6].

Very Recent study by J.-S. Zhou and J.B. Goodenough et al.[7] showed that *t-e* hybridization owing to the local site distortion and cooperative octahedral –site rotation is responsible for dramatic change in $T_N$ across the $RCrO_3$ family. They observed that the ferromagnetic coupling and antiferromagnetic cupling are comparable by means of *$t_3$-O-$e_0$* & *$t_3$-O-$t_3$* superexchange interaction, respectively. In other recent report, B. Rajeswaran and Khomskii et al.[8] showed the magnetically induced switchable polarization (P~ 0.2-0.8 μC/cm$^2$) in weak ferromagnetic $RCrO_3$ with magnetic rare-earth ion. Their work demonstrates the week ferromagnetism associated with the Cr-sublattice, play the crucial role in inducing the polar ordering and the drop in magnetization at spin–reorientation transition towards lower temperature, attributed to the change in Cr-spin configuration. Some of the earlier reports have demonstrated that compounds in $RCrO_3$ family showed temperature as well as field induced spin-reorientation transition quite below $T_N$ due to the change in spin configuration[6,9,10], where the magnetic symmetry analysis predicted the magnetoelectric (ME) effect below spin reorientation temperature in these compounds.[5]

In such systems, origin of ferroelctricity seems to be entirely different from conventional ferroelectrics[11], and can be explained in terms of improper character as observed in isostructural orthorhombic rare-earth manganite-family ($RMnO_3$).[12] In $RMnO_3$, improper ferroelectricity is associated to the spin-lattice interactions in a modulated off-centre symmetric magnetic structure[13,14], and the electric polarization arises from lattice distortions is explained in terms of the inverse Dzyaloshinski-Morya model.[15,16] Moreover, in some of the orhtomagnites, competition between ferromagnetic and antiferromagnetic exchanges lead to modulate spin–lattice coupling mediated ferroelectric ground states.[17] Therefore, the study of spin-phonon coupling is important in such systems to probe the local structural changes due to the complex magnetic- ordering, where the nearest neighbour (NN) and next nearest neighbour (NNN) exchange interactions are also important. In particular, Raman spectroscopy is proved to be a powerful technique to elucidate phonon renormalization induced by magnetic excitation where the observed phonon anomalies nearby magnetic transition temperature have been attributed to spin-phonon coupling.

In this paper, we report a detailed temperature dependent Raman spectroscopic-studies of one of the orthocromites $YCrO_3$(YCO), which has been reported as biferroic material with G-type antiferromagnetic ordering below 142K ($T_N$) and a ferroelectric transition ($T_C$) around 473K. Our study aimed to exploit the sensitivity of Raman spectroscopy to correlate the lattice dynamics and magnetic excitation states in this multifunctional material.

## 2. EXPERIMENTAL DETAILS:

Polycrystalline samples were prepared by solid state reaction method by taking the stoichiometric quantities of $Y_2O_3$ and $Cr_2O_3$ at 1550$^o$C for 15hours followed by several intermediate grinding and heating steps. Compositional analysis of the synthesized sample was carried out using Rigaku *ZSX* Primus-II X-Ray Fluorescence Spectrometer (Results are not shown). Phase purity was confirmed by Rietveld refinement of the x-ray powder diffraction data. Magnetic measurements as function of temperature in field cooled (FC) and zero field cooled (ZFC) modes, were carried out with quantum design SQUID magnetometer. A micro-Raman study was performed in the backscattering geometry by using a Jobin-Yvon T64000 triple spectrometer with grating (1800 grooves mm$^{-1}$). A diode pumped solid state ($\lambda$ =532nm) laser (Innova 90-5) was used as excitation source. The sample temperature was varied between 20 to 300K using closed cycle He-cryostats. It may be pointed out that in such experiments one has to be careful enough to choose the proper laser power. As high laser power can considerably affect the Raman spectra due to the local heating. We have carried out the Raman studies at low laser power in order to avoid the laser heating of the sample.

## 3. RESULTS AND ANALYSIS:

### 1. X-RAY DIFFRACTION ANALYSIS:

The structural characterization of the samples by X-ray diffraction is shown in Fig.1. The room temperature powder XRD-data were acquired within the *2Θ* range of 20–80$^o$.The spectral pattern has been indexed as orthorhombic perovskite (ICDD No.34-0365) with space group Pnma (62) and the peaks match suggest the single phase formation without secondary phase(s). The diffraction pattern was refined using Rietveld analysis to understand any subtle change in the structure. An excellent Rietveld fit profile was obtained without any change in the peak profiles or relative intensities, which rules out the existence of any anomalous structural change in centrosymmetric YCO. The refined lattice parameters and position of the constituent ions are listed in table1. The observed atomic positions and lattice parameters are in good agreement with those reported in literature.[18,19,20] Further, based on the formalism proposed by Y. Zhao et al.[21], the octahedral tilt angels are calculated using the observed atomic coordinates from X-ray refined data, as;

$$\theta = tan^{-1}\left[4\frac{\sqrt{W_{O1}^2 + U_{O1}^2}}{b}\right]$$

$$\phi = tan^{-1}\left[4\frac{\sqrt{W_{O2}^2 + U_{O2}^2}}{\sqrt{a^2 + c^2}}\right] \quad (1)$$

here, $W_{O1}, U_{O1}, W_{O2}$ and $U_{O2}$ are position coordinates of O1 and O2 atoms, respectively. The calculated values of tilt angles $\theta$[101] and $\phi$ [010] are found to be 17.01 and 12.24, respectively, which are in close agreement with values reported in Ref.4.

## 2. MAGNETIC PROPERTIES:

The magnetization as a function of temperature from 5K to 300K at magnetic field H = 100Oe is plotted in Fig. 2(a). Both ZFC and FC mode shows almost symmetrical changes in whole temperature range and the splitting around 142K confirmed the onset of the Cr-magnetic ordering from paramagnetic to the canted antiferromagnetic phase.[1,22] Below $T_N$, increase in magnetization with decrease in temperature and observed splitting in ZFC and FC modes at $T_N$ attributed to weak ferromagnetic ordering. The saturation in magnetization has been observed around ~60K. The negative value of the magnetization in ZFC mode could be understood in terms of competing WFM-FM interactions.[23] However, one of the artifact associated with the negative ZFC magnetization in ferro/ferri/antiferromagnetic materials, could be the small negative trapped field in the sample space and high coercive field, as observed in $CoCr_2O_4$.[24] To further understand the magnetic behavior, magnetization versus applied field curve (M-H) at different temperatures, 125K, 175K and 300K up to 20kOe, are shown in Fig.2(b). The hysteresis curve at 125K, measured after zero field cooling started from negative value of magnetization and this memory of negative magnetization of ZFC is not vanishes completely even by the maximum applied field of 2Tesla. The existence of clear opening of magnetization loop and lack of saturation magnetization attributed to canted antiferromagnetic ordering.[22] In such kind of magnetization loop the field dependence of the magnetic moment can be presented as: M(H) = $M_0$ + $\chi_{AFM}$.H;  where $\chi_{AFM}$ is the antiferromagnetic contribution and $M_0$ is the saturation moment of weak ferromagnetism.[19,25] It is cleared from the Fig.2(a) that below temperature $T$~60K the magnetization tends to saturate and the weak ferromagnetic saturation moment ~0.03$\mu_B$ is observed, which is similar to the behaviour observed by Tsushima et al.[26] for $ErCrO_3$. Above $T_N$, at 175K the M–H loop shows a straight line, which is expected for a paramagnetic material. A Curie-Weiss law fitting of inverse of the susceptibility ($\chi^{-1}$) versus temperature above $T_N$, was carried out as shown in inset of Fig.2 (a). Above $T$>150K, the data follow the Curie-Weiss behavior;

$$\chi^{-1} = \frac{(T - \theta_{CW})}{C} = \frac{3K_B}{N_A \mu_{eff}^2}(T - \theta_{CW}) \qquad (2)$$

here, $C$ is the Curie const, $N_A$ is the Avogadro's number, $\mu_{eff}$ is the effective magnetic moment, $\theta_{CW}$ is the Curie-Weiss temperature and $k_B$ is the Boltzmann constant. . In YCO, $Cr^{3+}$ ion has three localized 3d-electrons with possible spin quantum number S = 1/2 and 3/2. The possible $\mu_{eff}$ are 3.87$\mu_B$/f.u. for the high spin state (HS) and 1.73$\mu_B$/f.u. for the low spin state (LS) respectively. We observed the effective magnetic moment of 3.74$\mu_B$/f.u. from Curie-Weiss fitting, which is close to the expected theoretical value of 3.87$\mu_B$/f.u. This comparison of calculated value with theoretical one show $Cr^{3+}$ lies in HS in YCO. We also found Curie- Weiss temperature $\theta_{CW}$ = -520K from $\chi^{-1}$ vs $T$ plot, which is in agreement with antiferromagnetic interactions. Furthermore, the high coercive field (5.5kOe) and the remnant magnetization (0.53emu/gm) for YCO are indicative of a not so weak ferromagnetic component.

The above magnetization results demonstrate the coexistence of WFM and AFM ordering below $T_N$, whereas the unusual saturation in moment around 60K, could be the signature of change in magnetic symmetry[5,6,8,9,27], and merits further study. In order to get further insight regarding this complex magnetic behaviour and its correlation with lattice dynamic, we have performed temperature dependent Raman spectroscopic studies presented in the following section.

## 3. TEMPERATURE DEPENDENT RAMAN SPECTROSCOPY:

A normal mode decomposition obtained from a group theoretical analysis of *Pnma* symmetry of YCO, predicts a total of 60 normal modes ($\Gamma = 7A_g + 8A_{1u} + 5B_{1g} + 8B_{1u} + 7B_{2g} + 10B_{2u} + 5B_{3g} + 10B_{3u}$). Out of these 60 modes, 24 modes are Raman active and belong to the following irreducible representation: $7A_g + 5B_{1g} + 7 B_{2g} + 5 B_{3g}$.[28] Temperature dependent Raman measurements in the range of 20-300K, have been carried out on polished pallet of YCO in the spectral range of 100 to 600cm$^{-1}$. The representative unpolarized Raman spectra at various temperatures are shown in Fig.3 (a). The room temperature Raman spectra show 16 distinct Raman peaks occur at about 144, 166, 178, 213, 260, 274, 308, 335, 402, 419, 485, 495, 561, 644, 652 and 687 cm$^{-1}$. In the present study, the observed phonon modes were assigned in accordance with the previous reported results.[29, 30] Based on the symmetry assignment, Raman spectra in Fig.3 (a) confirmed the presence of seven $A_g$, three $B_{1g}$, three $B_{2g}$ and three $B_{3g}$ modes.[30] It may be noted that several materials undergo structural phase transition upon cooling and the Raman spectra of such materials show distinct change at the transition temperature. As can be seen from Fig.3 (a), the number of Raman peaks remains unchanged as the sample is cooled down to 20K, suggesting the absence of structural change across the magnetic transitions.

In $RCrO_3$, lattice dynamic calculation (LDC) predicts that the shapes of the higher Raman modes of each symmetry ($A_g$, $B_{1g}$, $B_{2g}$ and $B_{3g}$) involve pure or mixed *O1* and *O2* vibrations, whereas the lower Raman modes activated by mixed *R*-O1 vibrations either within the *XZ* planes ($A_g$ and $B_{2g}$ modes), or along the *Y*-axis ($B_{1g}$ or $B_{3g}$ modes).[31] Fig.4 shows the corresponding atomic displacement patterns of some selected Raman modes. It has been recently observed that two out of $7A_g$ modes are behaves as soft modes and their frequency scales linearly with the octahedral rotation angle.[4, 30]

Spin-phonon behaviour, which is the main objective here, is revealed through the temperature dependence of the line shape parameters in the temperature range of magnetic ordered phases. From this view point, the overall spectral range including the higher intensity modes from each symmetry, has been divided into two regions, i.e. 150-350 cm$^{-1}$ and 380 to 525cm$^{-1}$, respectively. Both the spectral ranges were fitted with the sum of Lorentzian line shapes as shown in Fig.3 (b) & (c). Temperature evolution of the extracted line shape parameters, i.e. peak positions and line width (FWHM) are plotted as function of temperature in Fig.5. Anomalous phonon behavior was noticed for the all studied modes. Upon cooling, all modes show conventional hardening down to $T_N$, coincident with the magnetic ordering temperature. Below $T_N$, the studied modes show a clear change in behaviour, with an anomalous softening or hardening on cooling down to $T^*$ ~60 K, where an additional phonon anomaly was identified. Below $T^*$, the modes show another change in behaviour, and this time all the modes get harder upon cooling. Here it may be emphasized that the observed hardening below $T^*$ does not appear to be a conventional phenomenon, since lattice contraction due to anharmonic effects are not expected to be significant in this temperature–range. Similarly, the temperature dependence of the phonon linewidths (FWHM) revealed clear anomaly at $T_N$ and $T^*$ for all studied modes.

**Discussion:**

As observed in the magnetization measurements, a mixed WFM and AFM phase has been observed below $T_N$. In such systems, the competitive interaction between Ferromagnetic and anti-ferromagnetic sate usually yield magnetic disorder, and therefore, even an arbitrarily small external applied field leads to misleading outcomes about the actual magnetic ground states.[17] From Currie –Weiss plot shown in Fig.2 (a), the value of I$\Theta_{CW}/T_N$I is quite high equal to 3. Based on earlier work by Neel[32] and Anderson[33], the Next nearest neighbour interaction (NNNI) is important to further understand the complex magnetic ordering in such antiferromagnetic compounds with high $\Theta_{CW}/T_N$ ratio. As we have already discussed the octrahedral tilting in YCO is strong enough to bend the Cr-O-Cr bond and consequently, the two facing oxygens are now enough close to each other, which facilitate the electrons to hop among next nearest neighbour (NNN) Cr-sites through Cr-O-O-Cr path. In such interaction, the hopping strength between the nearest-neighbour (NN) Cr-ions is reduced significantly.[34] Therefore, the NNNI along with NNI, will play an important role to stabilize the magnetic structure. In fact, the NNNI has been invoked to interpret the spiral and E-type magnetic structures in some of RMnO$_3$.[35, 36] Using the molecular field theory, the exchange constants can be related with the Currie-Weiss constant $\Theta_{CW}$ and Neel temperature $T_N$, as follows[27]:

$$J_1 = \frac{k_B(\Theta_{CW} - T_N)}{8s(s+1)} \ cm^{-1}$$

$$J_2 = \frac{1}{2} \cdot \frac{k_B(\Theta_{CW} + T_N)}{8s(s+1)} \ cm^{-1} \tag{3}$$

here, $J_1$ is the mean value of the two exchange constants for the six nearest neighbour Cr$^{3+}$-ions along the (X±Y) and Z axes, and $J_2$ is the mean value of the three exchange constants for the twelve next nearest neighbour Cr$^{3+}$ ions. (See Fig. 6)

We calculated exchange constants $J_1$= -15.34 cm$^{-1}$ and $J_2$= - 4.54 cm$^{-1}$, by using the values of $\Theta_{CW}$ and $T_N$. It is worthwhile to note that the value of $J_2$, which is corresponding to the average of NNNI is not small and nearly four times than the previously measured value of 1 cm$^{-1}$ by Tsushima et al.[27] in YCO. In the same paper, Tsushima studied the magnetic field and temperature dependence of absorption spectra, where the temperature evolution of the integrated intensity of observed sidebands (arises due to the combined excitation of a Cr-exciton and a Cr-magnon at the Brillouin Zone boundry) showed a sudden lowering in band intensity below $T\sim$60K, plausibly due to the change in spin-anisotropic axis, thus yields different spin configuration below 60K, as observed latter on in some other RCrO$_3$.[37] Moreover, in a recent paper Duran et al.[22] predicted a spin reorientation transition in YCO, based on specific heat measurements. They observed a broad peak around 60K in $C_{mag}/T$ plot. Such large residual weight of magnetic contribution in total specific heat ($C_P$) again indicates the change in spin dynamics across $T\sim$60K, form its ground state. These findings are in agreement with our magnetization and Raman spectroscopic results, where the unusual change in magnetization and phonons line-shape behaviour has been found in the vicinity of spin reorientation temperature at $T\sim$ 60K.

Furthermore, it has been observed from Raman results that the temperature evolution of some of the phonon modes show softening, whereas some of the modes show hardening as temperature decreases below $T_N$. In order to calculate the effect of the magnetic exchange interactions on the phononic behaviour, we can describe the purely anharmonic temperature dependence of the frequency of the different modes by the model[38]:

$$\omega(T) = \omega_0 + A\left[1 - \frac{2}{e^x - 1}\right] \qquad (4)$$

here, $\omega(T)$ is the phonon frequency at $T$(K), $x = \hbar/2kT$, with thermal energy $kT$. The parameters $\omega_0$ and $A$ are the adjustable parameters, which were optimized for fitting the experimental data for $T>T_N$. In this view point, the some of the modes were fitted using the above equation, as shown in Fig.7, where the solid red line represents the best fit to the Eq.4 in the temperature range above $T\sim$140K. In the temperature range below $T\sim$140 K, the extrapolated red line showed a significant deviation from the behaviour predicted by Eq. 4, and softening as well as hardening of the phonon frequencies is clearly visible below $T_N$. Whereas, a sudden increase in phonon frequency observed below $T^*$. According to the Fig.7, the phonon anomalies are in better agreement with the magnetization behaviour in terms of derivative of the magnetic susceptibilities. This correlation of anomalous phonon behaviour with magnetic excitations is attributed to spin-phonon coupling caused by the phonon modulation of spin exchange integral.[39] Moreover, the increase in linewidth (FWHM) of some phonon modes (345, 281 cm$^{-1}$) below $T_N$ and $T^*$ indicates a decrease in the lifetime of the phonons due to spin-phonon coupling.[40]

Furthermore, the shift in frequency of a given phonon mode as a function of temperature, due to spin-phonon coupling is determined by the following spin-spin correlation function[41, 42];

$$\Delta\omega = \omega - \omega_0 = \lambda <\vec{s_i}.\vec{s_j}> \qquad (5)$$

here, $\omega$ denotes the renormalized phonon frequency at a fixed temperature, $\omega_0$ is the frequency in the absence of spin-phonon coupling, and $\lambda$ is the spin-phonon coupling constant. In our case, the values of $\Delta\omega$ are found to be positive as well as negative for different modes (see Fig.7). This kind of positive and negative shifting has been recently observed in doped orthorhombic $Eu_{1-x}Y_xMnO3$[12, 17], due to presence of ferromagnetic and anitferromagnetic competitive interactions. Our magnetization results also confirmed the presence of weak ferromagnetoic as well as antiferromagnetic interaction in YCO below $T_N$. In such case, the $\omega - \omega_0$ frequency shift can be interpreted using the theoretical model represented by the following equation[42, 43];

$$\Delta\omega = \omega - \omega_0 \propto -k_1 <\vec{s_i}.\vec{s_j}> + k_2 <\vec{s_i}.\vec{s_k}> \qquad (6)$$

here, *k1* and *k2* are the spin dependent force constants of the lattice vibrations deduced as the second derivatives of the exchange integrals with respect to the phonon displacement.[12] This model predicts negative or positive shifts of $\Delta\omega$, depending on the relative strength between the ferromagnetic and antiferromagnetic exchange interactions, associated with the phonon mode being considered.[12] However, Moriera et al.[12] reported that the WFM character of the compound cannot be explained by this model, because it is not associated with the second derivatives to the phonon displacements, represented by the *k*1 and *k*2 coefficients. From magnetization results, the presence of high coercive field and the remnant magnetization indicates not so weak ferromagnetic component in YCO. Hence, one could explain the positive or negative shift in phonon frequency as a consequence of competitive WFM and AFM exchanges in YCO.

Beside this, in order to further verify the spin-phonon coupling, we utilized the mechanism proposed by Granado et al.[39], which suggest the effect of phonon renormalization below $T_N$ is proportional to the nearest neighbour spin-spin correlation function $<\vec{s_i}.\vec{s_j}>$ , and under the molecular filed approximation this phonon renormalization scales as the square of the sublattice magnetization:

$$<\vec{s_i}\cdot\vec{s_j}> \approx 4\ \{M_{sublat}(T)/4\mu B\}^2 \qquad (7)$$

and, the following equation is used to fit the phononic behaviour[39];

$$\Delta\omega \approx -\frac{2}{m\omega}\frac{\partial^2 J}{\partial u^2}\ \{M_{sublat}(T)/4\mu B\}^2 \qquad (8)$$

Before moving further, it is worthwhile to note that, in compounds which shows complex magnetic behaviour ascribed by mixed FM and AFM phases, the onset of phonon renormalization could be locate at temperatures well above $T_N$, and therefore, it required certain modification in above mechanism, as proposed by Laverdière et al.[44] In our case, the phonon renormalization started right below the ordering temperature $T_N$, which indicates the temperature behaviour of phonon modes below $T_N$ could be explain fairly well using above model.

According to Fig.8, softening of the mode 485 cm$^{-1}$ can be fitted quite well using Eq.8. Although, the unusual phonon behaviour below $T^*$, could not be scaled by the same equation, which might indicates some different mechanism of interaction between spin and phonon systems below $T^*$ and required further consideration.

**Conclusions:**

Temperature dependent lineshape analysis of selected phonon modes in YCO were investigated by means of Raman scattering, complemented by magnetization measurements. A mixed WFM+AFM phase has been observed below $T_N$. Raman and magnetization results reveal the new characteristic temperature $T^*$ , plausibly due to the change in $Cr^{3+}$-spin configuration, below which magnetization get saturated and the phonons frequency shifted anomalously .

The sign of phonon shift below the magnetic ordering temperature in the virtue of mixed WFM and AFM exchange interaction demonstrate the strong correlation of phonons with magnetic excitations in YCO. Our study could be useful to further understand the complex magnetic behaviour and spin-phonon interactions in some of the multiferroic rare-earth chromites, in which the spin-reorientation and/or magnetization reversal has been found below magnetic ordering temperatures.

**ACKNOWLEDGEMENT:** The authors acknowledge financial support from DOE (grant DE-FG02-ER46526).

**FIGURE CAPTIONS:**

- Fig.1: Rietveld refined XRD pattern of bulk single phase $YCrO_3$ at 300K.
- Fig.2: (a) Magnetization versus temperature plot in ZFC and FC mode in the range of 5-300K. The inset shows the inverse susceptibility versus temperature plot. The blue line shows Curie-Weiss law fitting at $T>150K$ to 300K. (b) Magnetic hysteresis loops at 125K, 175K and 300K.
- Fig.3: (a) The temperature dependent Raman spectra at selected temperature between 20- 300K, (b) & (c) represent the two selected spectral ranges fitted with the sum of Lorentzian line shapes.
- Fig.4 Schematic atomic-displacement patterns for some of the Raman active modes experimentally observed in Fig.3.
- Fig.5 Temperature dependence of the line shape parameters (peak positions and FWHM) of some selected Raman modes.
- Fig.6 Schematic representation of two exchange interactions in unit cell of $YCrO_3$.
- Fig.7 Comparison of the temperature dependence of Raman peak positions (selected modes) with those of the derivative of magnetic susceptibility $\left(\frac{d\chi}{dT}\right)$ in FC and ZFC mode. The fitted solid red lines to the Raman peak positions have been obtained from the best of Eq. 4.
- Fig.8 Temperature versus frequency shift of $B_{3g}$ (3) [485cm$^{-1}$] mode with respect to its value at $T$=155K. The square of sublattice magnetization $[M_{sublat}(T)/4\mu B]^2$ has been represented by solid red line.

**TABLE:**

- Table 1. Refined structural parameters and atomic positions as obtained from powder XRD-data for $YCrO_3$; [Space group Pnma: Y: 4c (x,0.25,z); Cr: 4b (0,0,0.5); O(1): 4c (x, 0.25, z) and O(2): 8d (x,y,z)].

| T = 300K | | | |
|---|---|---|---|
| $\chi^2$= 1.93, $R_p$=7.4, $R_{wp}$= 10.3, $R_{exp}$=7.42<br>Density = 5.762 g/cm$^3$,<br>a= 5.5122(4), b= 7.5349(2), c= 5.2461(3) | | | |
| Atomic positions | x | y | z |
| Y(4c) | 0.0655(2) | 0.25 | -0.0163(4) |
| Cr(4b) | 0 | 0 | 0.5 |
| O1(4c) | 0.4646(3) | 0.25 | 0.1118(1) |
| O2(8d) | 0.3018(1) | 0.0532(3) | -0.3258(2) |

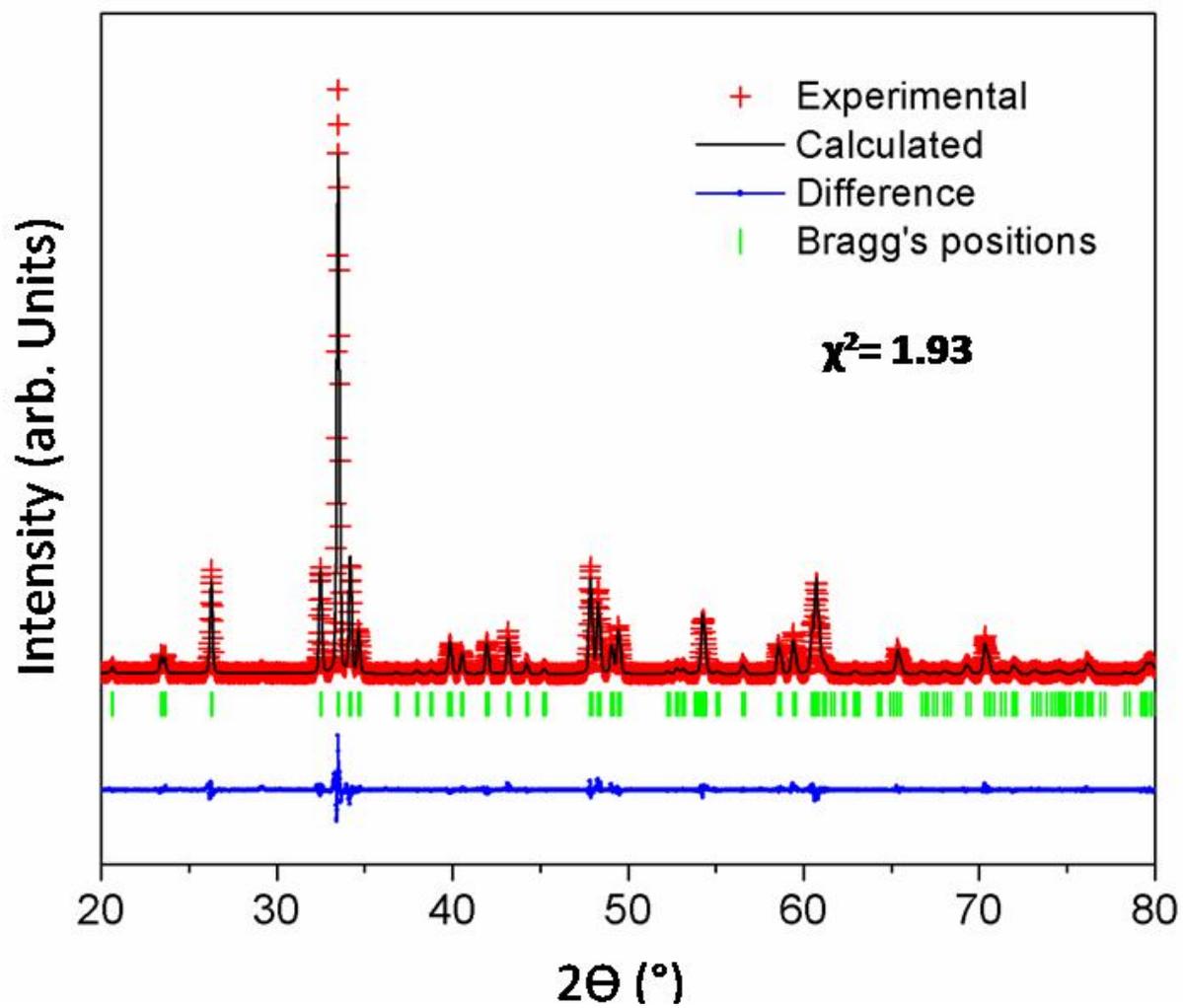

Fig.1 Sharma et al.

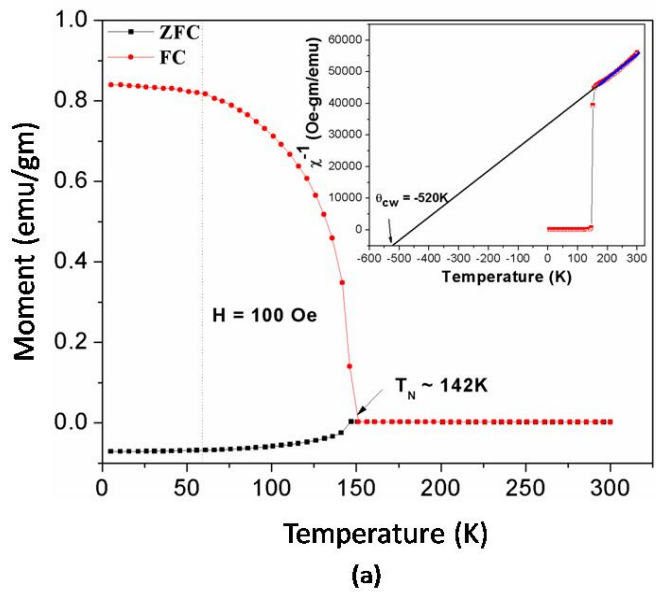 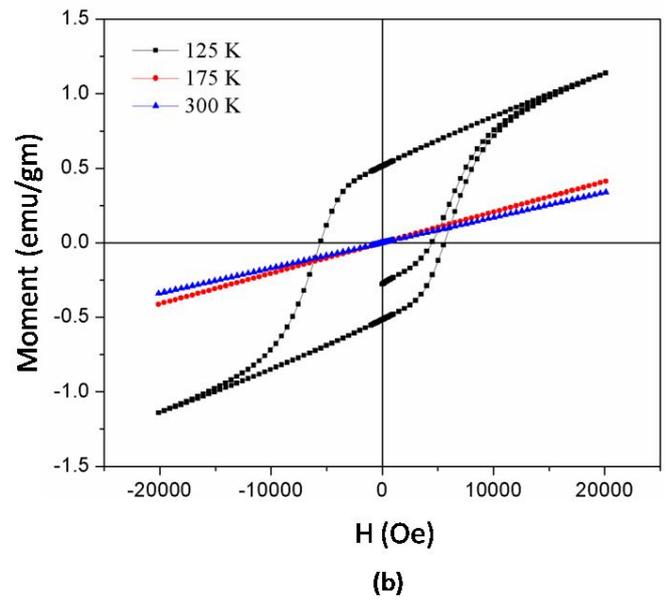

**Fig.2 Sharma et al.**

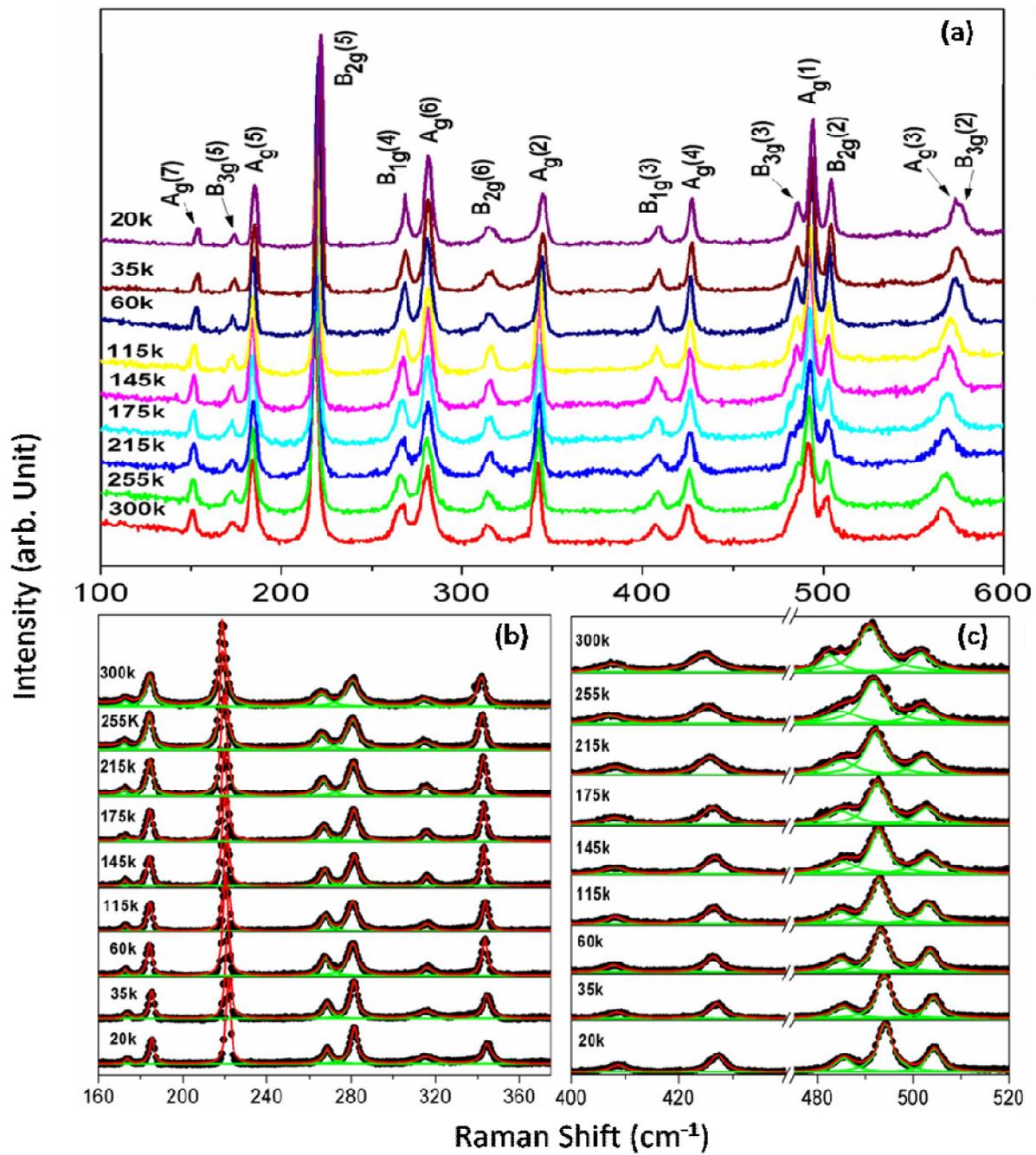

Fig.3 Sharma et al.

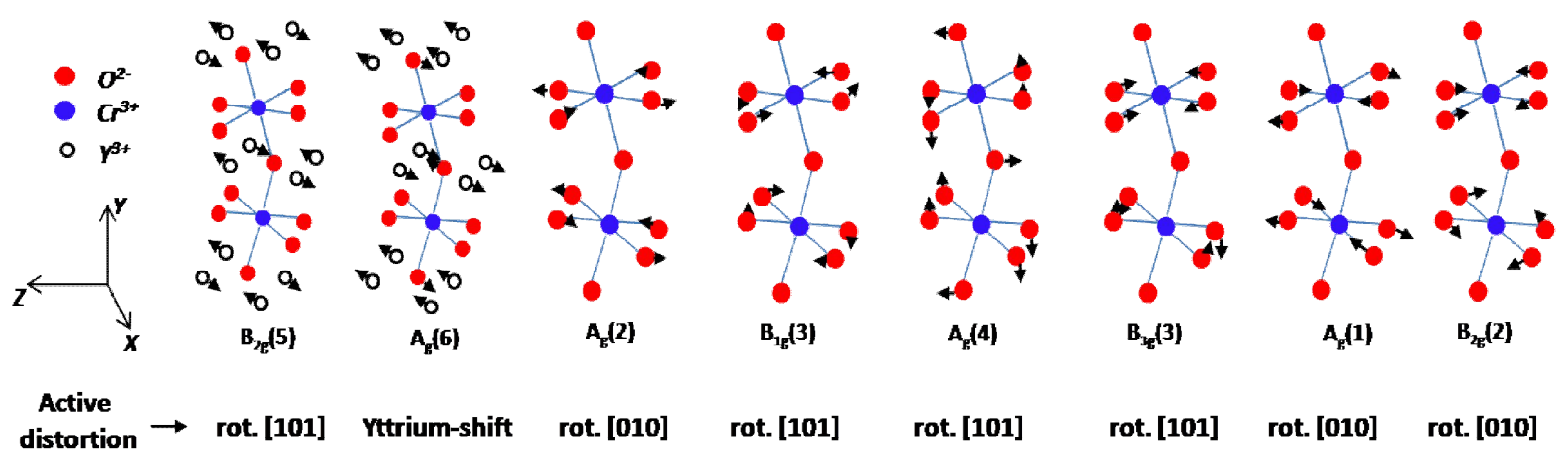

Fig.4 Sharma et al.

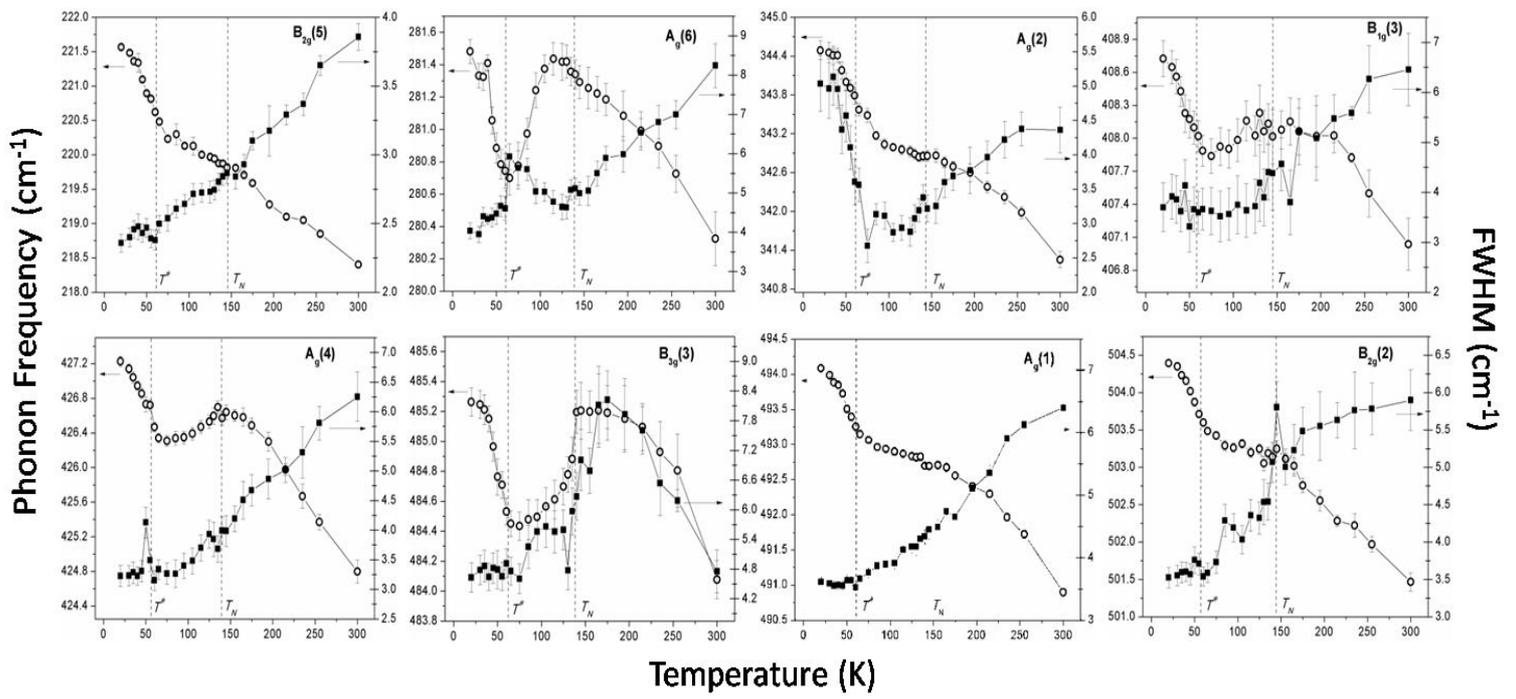

Fig.5 Sharma et al.

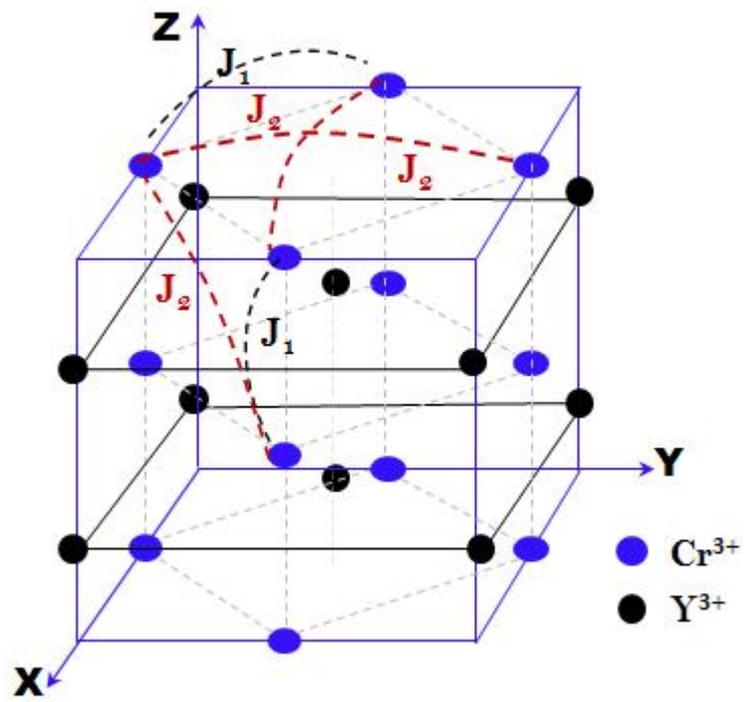

**Fig.6 Sharma et al.**

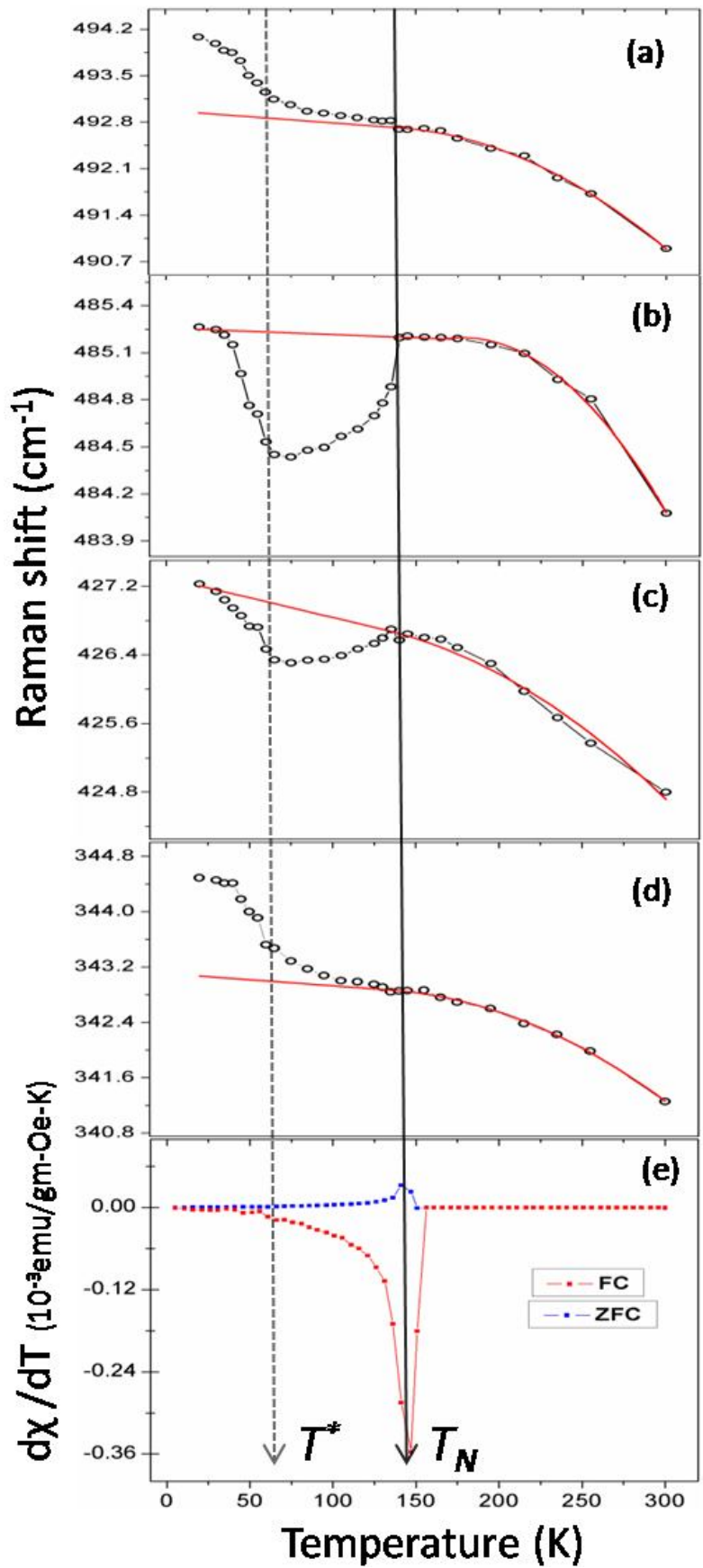

Fig.7 Sharma et al.

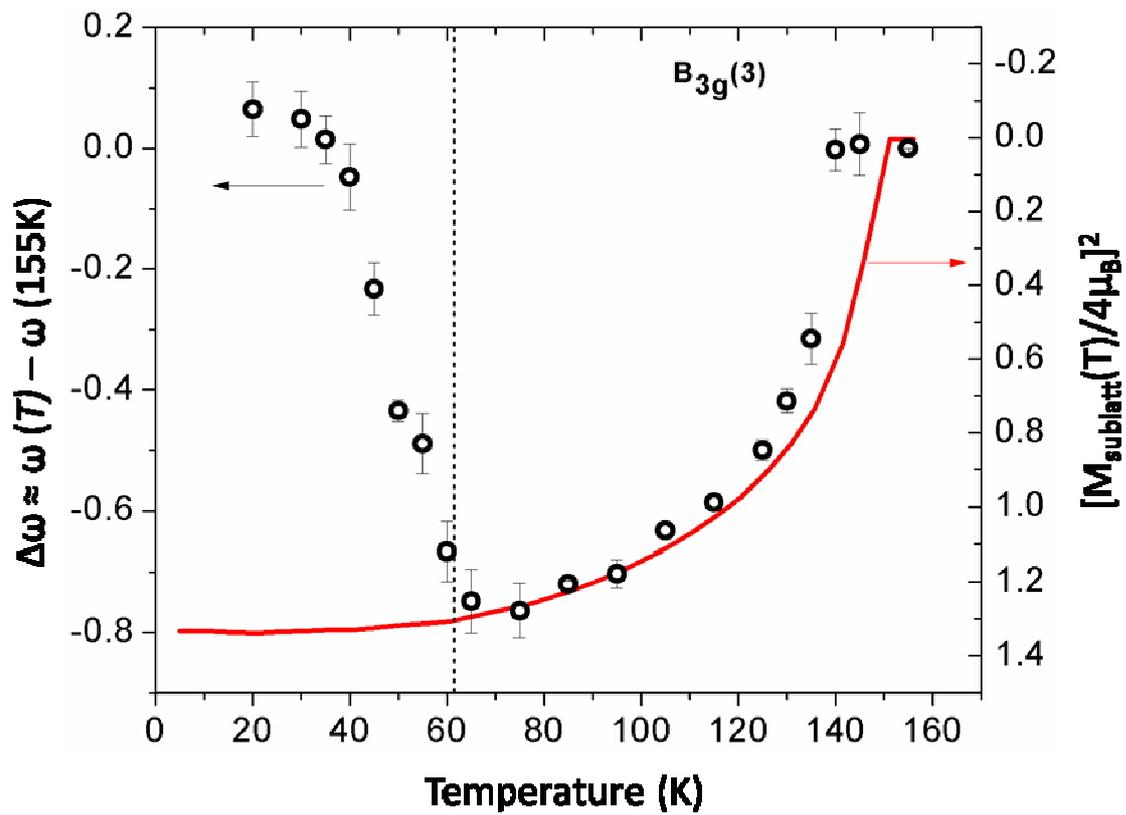

Fig.8 Sharma et al.